\documentclass[conference]{IEEEtran}
\usepackage{amssymb}
\usepackage{mathrsfs}
\usepackage{amsmath}
\usepackage{latexsym}
\usepackage{graphicx}
\usepackage{bbding}
\usepackage{indentfirst}
\IEEEoverridecommandlockouts

\begin{document}
\title{Energy Efficient Joint Source and Channel Sensing in Cognitive Radio Sensor Networks}
\author{\authorblockN{Huazi~Zhang$^{1,2}$, Zhaoyang~Zhang$^{1,2}$, Xiaoming~Chen$^{1,2}$, Rui~Yin$^{1,2}$ \\1. Institute of information and
communication engineering, Zhejiang University, Hangzhou 310027,
China.
\\2. Zhejiang provincial key laboratory of information network
technology, Hangzhou 310027, China.
\\Email: ning\_ming@zju.edu.cn
\thanks{This work was supported by the National Hi-Tech Research
and Development Program (863 program) of China (No. 2007AA01Z257),
National Basic Research Program (973 Program) of China (No.
2009CB320405) and National Natural Science Foundation Program of
China (No. 60802012, 60972057).}}} \maketitle

\begin{abstract}
A novel concept of Joint Source and Channel Sensing (JSCS) is
introduced in the context of Cognitive Radio Sensor Networks (CRSN).
Every sensor node has two basic tasks: application-oriented source
sensing and ambient-oriented channel sensing. The former is to
collect the application-specific source information and deliver it
to the access point within some limit of distortion, while the
latter is to find the vacant channels and provide spectrum access
opportunities for the sensed source information. With in-depth
exploration, we find that these two tasks are actually interrelated
when taking into account the energy constraints. The main focus of
this paper is to minimize the total power consumed by these two
tasks while bounding the distortion of the application-specific
source information. Firstly, we present a specific slotted sensing
and transmission scheme, and establish the multi-task power
consumption model. Secondly, we jointly analyze the interplay
between these two sensing tasks, and then propose a proper sensing
and power allocation scheme to minimize the total power consumption.
Finally, simulation results are given to validate the proposed
scheme.
\end{abstract}

\section{Introduction}
Wireless Sensor Networks (WSN) are capable of monitoring physical or
environmental information (e.g. temperature, sound, pressure), and
collecting them to certain access points according to various
applications. The extensive deployment of WSN has changed our lives
dramatically. However, current WSN nodes usually operate on
license-exempt Industrial, Scientific and Medical (ISM) frequency
bands \cite{CogSeNet}, and these bands are shared with many other
successful systems such as Wi-Fi and Bluetooth, causing severe
spectrum scarcity problems \cite{CWSN}. To deal with such problems,
a new sensor networking paradigm of Cognitive Radio Sensor Network
(CRSN) which incorporates cognitive radio capability on the basis of
traditional wireless sensor networks was introduced \cite{CRSN}.
CRSN nodes operate on licensed bands and can periodically sense the
spectrum, determine the vacant channels, and use them to report the
collected source information. The main design principles and
features of CRSNs are discussed openly in literature
\cite{CogSeNet}-\cite{CWSNsurvey}. According to these literatures,
CRSN enjoys many advantages, such as efficient spectrum usage,
flexible deployment and good radio propagation property. However,
WSN nodes are low cost and usually equipped with a limited energy
source, such as a battery, and CRSN nodes also inherit this
fundamental limitation. What¡¯s more, the CRSN node bears one more
task of spectrum sensing, and this task also consumes energy. This
fact makes the energy scarcity problem in CRSN even more severe.
Hence, how to minimize the total energy consumption for CRSN node
and thus make the system the most energy efficient has become an
urgent problem.

In our viewpoint, there are two basic types of sensing tasks for the
CRSN node, one is Application-Oriented Source Sensing (AppOS) and
the other is Ambient-Oriented Channel Sensing (AmOS). By source
sensing we mean the process of collecting source information (e.g.
temperature, sound) and delivering it to the Access Point (AP), and
by channel sensing we mean the process of periodically sensing the
ambient radio environment and determining the vacant channels for
opportunistic spectrum access.

The energy saving problems of both AppOS and AmOS have been
investigated separately in existing literature. The energy
consumption models for AppOS have been established in the context of
conventional WSN. The energy-distortion tradeoffs in
energy-constrained sensor networks is investigated in \cite{E-D1},
and energy efficient lossy transmission for wireless sensor networks
is studied in \cite{E-D3}, for Gaussian sources and unlimited
bandwidth. Another issue of energy efficient AmOS has also been
studied separately, in the cognitive radio scenario. Maleki has
designed a sleep/censor scheme to reduce spectrum sensing energy
\cite{E-CR2}. Su and Zhang proposed an energy saving spectrum
sensing scheme by adaptively adjusting the spectrum sensing periods
utilizing PU's activity patterns \cite{E-CR3}. \cite{E-CR5} studied
the influence of sensing time on the probability of detection and
probability of false alarm.

However, the unique mechanism of CRSN is that every CRSN node
performs AppOS and AmOS at the same time. This requires
us to consider the resource saving problems of AppOS and
AmOS jointly. In order to prolong the lifespan, there is a
need to properly distribute limited power into these two concurrent tasks. On one hand, if we
put excessive power into AppOS,
the resources left for AmOS will be diminished. We can obtain
more precise and unaffected application-specific source information.
But, due to the lack of channel resource information, acquired source information can not
be delivered timely and effectively to AP. Furthermore, the probability of miss detection of Primary
Signals can be prominently high. It will cause interference
to the underlying Primary System. On the other hand, if we put too much power into AmOS, we can obtain enough reliable
channel access opportunities, and reduce the interference
to the Primary System. Vice versa, the power left for AppOS
is not enough for delivering source information at a coding
rate capable of meeting the distortion requirement despite the implementing of distributed source coding in the sensor
network. Therefore, our paper mainly aims at tackling this joint energy
saving problem, which has not been considered before. The main
contributions of our work are as follows: we jointly model the power
consumption of AmOS and AppOS and use the transmission probability
to bond these two interrelated tasks; we find that within bounded
distortion, there is always a minimal total power consumption and
corresponding power allocation scheme for the CRSN system, which is
the most power efficient solution.

The rest of this paper is organized as follows. In Section II, we
make basic assumptions about the CRSN, and provide a brief
introduction of the considered system. In Section III, we give
detailed models and jointly analyze the power consumption of AppOS
and AmOS. Then, several simulation results are presented in Section
IV to further validate our analysis. Finally, the whole paper is
concluded in Section V.

\section{System Model}
\begin{figure}[h] \centering
\includegraphics [width=0.50\textwidth] {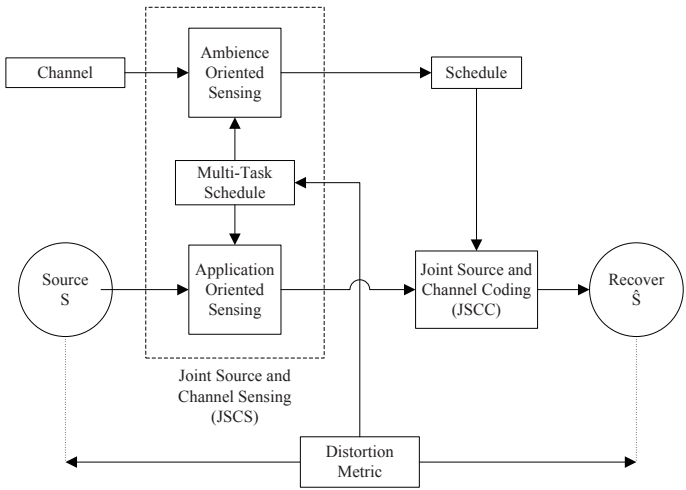}
\caption {An overview of the considered system model.}
\label{Fig1}
\end{figure}
In this paper, we consider the Multi-task Sensing architecture in
CRSN nodes, as shown in Fig.\ref{Fig1}. We observe two dominating
features of the considered CRSN:

\textbf{Feature 1}: In a CRSN node, two major tasks need to be
modeled as follows:

\begin{enumerate}

\item Application Oriented Source Sensing (AppOS)

We define source sensing as the
process of collecting various source information (e.g. temperature,
pressure, position, etc.) according to the application-specific
demand and delivering it to the Access Point (AP).
The main objective of AppOS is realizing accurate acquisition
of the source information.

\item Ambient Oriented Channel Sensing (AmOS)

Channel sensing is the process
of periodically sensing the ambient radio environment by
means of spectrum sensing and energy detection. It, thus,
determines the vacant channels for opportunistic spectrum
access or perceives energy distribution of surrounding nodes
for cooperation. The main objective of AmOS is to realize
effective and efficient exploration of spectrum resources.

\end{enumerate}

\textbf{Feature 2}: As a characteristic inherited from the
traditional WSN, every CRSN node is power-constrained due to limited
energy supply. Both AmOS and AppOS consume energy. We have to save
as much energy as possible while delivering the source information
to AP within bounded distortion.

\begin{figure}[h] \centering
\includegraphics [width=0.30\textwidth] {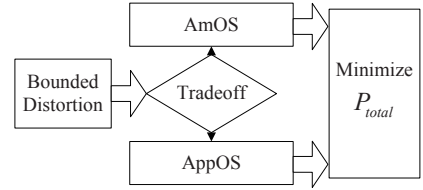}
\caption {The interplay between the two tasks.} \label{Fig2}
\end{figure}

Fig.\ref{Fig2} depicts the subtle interplay between AppOS and AmOS sensing under the interference, distortion and
power resource constraint. It's like you have two ears
listening to two distinct but related objects in a noisy
environment. On the one hand, your left ear listens to the monitored
source, trying to hear the most undistorted sound. On the other
hand, your right ear listens to the slight ambient sound on the
radio spectrum, because you are not allowed to speak when others
talk. Our goal is to optimally balance the two ears and make them
the most efficient.

\subsection{Slotted Sensing and Transmission Scheme}

We present a specific sensing and transmission scheme below in
Fig.\ref{Fig3}:

\begin{figure}[h] \centering
\includegraphics [width=0.45\textwidth] {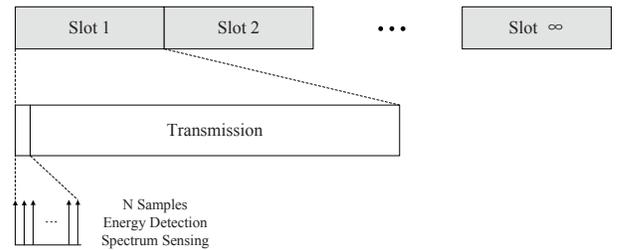}
\caption {Slotted sensing and transmission scheme.} \label{Fig3}
\end{figure}

Every node partitions the time domain into periods, namely slots, of
equal length $T$. At the beginning of each slot, the cognitive
sensor node makes a decision on whether or not to transmit based on
the $N$ samples energy detection spectrum sensing result. The
spectrum sensing is always performed ahead of the data transmission.
In our scheme, we should point out three basic assumptions:

\textbf{AS1:} Both spectrum sensing and data transmission consume
energy. And the total energy is limited in one CRSN node.

\textbf{AS2:} The slot length $T$ is short enough, so that the
status of primary user activity remains the same during one slot.

\textbf{AS3:} The time period of spectrum sensing is rather short
compared with transmission period and thus can be omitted.

In the following sections, we will establish detailed models for
both sensing tasks and analyze the power consumption tradeoff
between them.

\section{Energy Efficient Joint Source and Channel Sensing}

In this section, the relationships between power consumption and
performances are discussed. We present specific models for both AmOS
and AppOS, and then jointly analyze the relationship and tradeoff
between them. We prove that optimal power allocation scheme can
indeed be obtained.

\subsection{AmOS: Energy Detection based Spectrum Sensing}

In recent years, many methods have been developed for spectrum
sensing, including matched filter detection, energy detection and
cyclostationary feature detection. Among them, energy detection is
the most popular spectrum sensing scheme. It is the most suitable
for CRSN node due to its simplicity of hardware implementation and
low signal processing cost. Therefore, we choose energy detection as
our spectrum sensing technique for CRSN node.

We assume that the CRSN node operates at certain carrier frequency
$f_c$ with bandwidth $W$, and samples the signal within this range
$N$ times per slot.

The discrete signal that the CRSN node receives can be represented as:
\begin{equation}
y\left( n \right) = \left\{ {\begin{array}{*{20}c}
   {s\left( n \right) + u\left( n \right),\quad {\cal H}_1 :{\rm{primary\; user\; is\; active}\quad\quad}}  \\
   {u\left( n \right),\quad\quad\quad{\cal H}_0 :{\rm{primary\; user\; is\; inactive}}}  \\
\end{array}} \right.
\end{equation}

The primary signal $s(n)$ is independent, identically distributed
(i.i.d) random process with zero mean and variance $E\left[ {\left|
{s(n)} \right|^2 } \right] = \sigma _s^2$, and the noise $u(n)$ is
i.i.d random process with zero mean and variance $E\left[ {\left|
{u(n)} \right|^2 } \right] = \sigma _u^2$. We assume that the
primary signal $s(n)$ is MPSK complex signal, and the noise $s(n)$
is complex Gaussian.

As the performance criteria for the proposed spectrum sensing
method, the two important parameters worth mentioning are:
probability of detection and probability of false alarm. The
probability of detection, denoted as $P_D$, is the probability that
the CRSN node successfully detects the primary user when it's
active, under hypothesis $ \mathcal {H}_1$. The probability of false
alarm, denoted as $P_{FA}$, is the probability that the CRSN node
falsely determines the presence of primary signal when the primary
user is actually inactive, under hypothesis $\mathcal {H}_0$.

The energy detector is as follows:
\begin{equation}
T(y) = \frac{1}{N}\sum\limits_{n = 1}^N {\left| {y(n)} \right|^2
}\label{eqn3}
\end{equation}

According to the Central Limit Theorem, the statistics $T(y)$ is
approximately Gaussian distributed when $N$ is large enough under
both hypothesis $\mathcal {H}_1$ and $\mathcal {H}_0$. The
probability density function(PDF) of statistics $T(y)$ can be
expressed as:
\begin{equation}
T(y) \sim \left\{ {\begin{array}{*{20}c}
   {\mathcal {N}\left( {\mu _0 ,\sigma _0^2 } \right),{\rm{\ \ \ under \ \ \mathcal {H}_0}}}  \\
   {\mathcal {N}\left( {\mu _1 ,\sigma _1^2 } \right),{\rm{\ \ \ under \ \ \mathcal {H}_1}}}  \\
\end{array}} \right. \label{eqn4}
\end{equation}

When the primary signal $s(n)$ is MPSK complex signal, and the noise
$s(n)$ is complex Gaussian \cite{E-CR5}, %we have $\mu _0  = \sigma
%_u^2$ and
%\begin{equation}
%\begin{array}{l}
%\sigma _0^2  = \frac{1}{N}\left[ {E\left| {u(n)} \right|^{^4 }  -
%\sigma _u^4 } \right] = \frac{1}{N}\sigma _u^4
% \end{array} \label{eqn5}
%\end{equation}
%while $\mu _1  = \sigma _s^2  + \sigma _u^2$ and
%\begin{equation}
%\begin{array}{l}
% \sigma _1^2  = \frac{1}{N}\left[ {E\left| {s(n)} \right|^{^4 }  + E\left| {u(n)} \right|^{^4 }  - \left( {\sigma _s^2  - \sigma _u^2 } \right)^2 } \right] \\
% {\rm{    }} = \frac{1}{N}\left( {2\sigma _s^2 \sigma _u^2  + \sigma _u^4 } \right) \\
% \end{array} \label{eqn6}
%\end{equation}
%
%By setting the detection threshold $\varepsilon$,
we can derive the
probability of false alarm:
\begin{equation}
P_{FA} \left( {\varepsilon ,N} \right) = Q\left( {(\frac{\varepsilon
}{{\sigma _u^2 }} - 1)\sqrt N } \right) \label{eqn7}
\end{equation}
where $Q\left( x \right) = \frac{1}{{2\pi }}\int_x^\infty  {\exp ( -
\frac{{t^2 }}{2})dt}$ is the tail probability of the standard normal
distribution (also known as the Q function).

For certain threshold $\varepsilon$, the probability of detection
can be expressed as:
\begin{equation}
P_D \left( {\varepsilon ,N} \right) = Q\left( {(\frac{{\varepsilon -
\sigma _s^2 }}{{\sigma _u^2 }} - 1)\sqrt {\frac{N}{{2\sigma _s^2
/\sigma _u^2  + 1}}} } \right)\label{eqn8}
\end{equation}

Since the slot period $T$ is short enough, we can assume that the
primary user activity keeps unchanged during a single slot.

When the CRSN node fails to detect the PU signal, its signal will
collide with the primary user signal and bring interference into the
PU system. We denote $P_E=1- P_D$ as the probability of missed
detection, and have the following assumption:

\textbf{AS4:} There is a maximal missed detection probability that
the PU system can tolerate, and a typical value for this parameter
is 0.1 \cite{E-CR6}. $P_E$ should be smaller than this value.

%We calculate $\varepsilon$ corresponding to certain $P_E$:
%\begin{equation}
%\varepsilon  = \frac{{Q^{ - 1} \left( {1 - P_E } \right)}}{{\sqrt
%{\frac{N}{{2\sigma _s^2 \sigma _u^2  + \sigma _u^4 }}} }} + \sigma
%_s^2  + \sigma _u^2 \label{eqn9}
%\end{equation}

Because $Q\left(  \cdot  \right)$ is monotonically decreasing, we
find that the probability of false alarm drops as the sample number
$N$ increases:
\begin{equation}
P_{FA}  = Q\left( {\sqrt {2\sigma _s^2 /\sigma _u^2  + 1} Q^{ - 1}
\left( {1 - P_E } \right) + \sqrt N \sigma _s^2 /\sigma _u^2 }
\right) \label{eqn10}
\end{equation}

The resulted probability that the CRSN node is allowed to transmit
is:
\begin{equation}
\begin{array}{*{20}c}
   {p'_t  = \left( {1 - P_{FA} } \right)p({\rm{H}}_0 ) + \left( {1 - P_D } \right)p({\rm{H}}_1 )}  \\
   {\begin{array}{*{20}c}
   { = \left( {1 - Q\left( {\sqrt {\frac{{2\sigma _s^2 }}{{\sigma _u^2 }} + 1} Q^{ - 1} \left( {1 - P_E } \right) + \sqrt N \frac{{\sigma _s^2 }}{{\sigma _u^2 }}} \right)} \right)}  \\
   { \times p({\rm{H}}_0 ) + P_E p({\rm{H}}_1 )}  \\
\end{array}}  \\
\end{array} \label{eqn11}
\end{equation}
where $p({\rm{H}}_0 )$ and $p({\rm{H}}_1 )$ are the inactive and
active probabilities of the primary user, respectively. Leaving out
the collision probability $P_C = P_E p({\rm H}_1 )$, we can obtain
the effective transmission probability available for CRSN node:
\begin{equation}
\begin{array}{l}
 p_t  = p'_t  - P_E p({\rm{H}}_1 ) \\
  = \left( {1 - Q\left( {\sqrt {\frac{{2\sigma _s^2 }}{{\sigma _u^2 }} + 1} Q^{ - 1} \left( {1 - P_E } \right) + \sqrt N \frac{{\sigma _s^2 }}{{\sigma _u^2 }}} \right)} \right)p({\rm{H}}_0 ) \\
 \end{array} \label{eqn12}
\end{equation}

Denoting the energy consumed in one sample as $E_{sample}$, the
average AmOS power consumption can be expressed as:
\begin{equation}
P_{_{AmOS}}  = \frac{{E_{sample}  \times N}}{T} \label{eqn13}
\end{equation}

Rewriting the AmOS power expression with respect to the effective
transmission probability $p_t$ gives:
\begin{equation}
\begin{array}{l}
 P_{_{AmOS}} \left( {p_t } \right) \\
  = \left( {\frac{{Q^{ - 1} \left( {1 - \frac{{p_t }}{{p({\rm{H}}_0 )}}} \right) - \sqrt {\frac{{2\sigma _s^2 }}{{\sigma _u^2 }} + 1} Q^{ - 1} \left( {1 - P_E } \right)}}{{\sigma _s^2 /\sigma _u^2 }}} \right)^2  \times \frac{{E_{sample} }}{T} \\
 \end{array} \label{eqn14}
\end{equation}

Note that (10) is valid only when $p_t$ falls in the range of:
\begin{equation}
\left( {1 - Q\left( {\sqrt {\frac{{2\sigma _s^2 }}{{\sigma _u^2 }} +
1} Q^{ - 1} \left( {1 - P_E } \right)} \right)} \right)p({\rm{H}}_0
) < p_t  < p({\rm{H}}_0 ) \label{eqn15}
\end{equation}

When $0 < p_t  < \left( {1 - Q\left( {\sqrt {\frac{{2\sigma _s^2
}}{{\sigma _u^2 }} + 1} Q^{ - 1} \left( {1 - P_E } \right)} \right)}
\right)p({\rm{H}}_0 ) $, the transmission probability is so small
that the requirement of \textbf{AS4} can always be met. In this
case, we don't have to do any spectrum sensing, and $ P_{_{AmOS}}
\left( {p_t } \right) = 0$.

\subsection{AppOS: Distortion-Constrained Source Sensing}

In this subsection, we step forward to explore the connection
between $p_t$ and average AppOS power $P_{_{AppOS}}$. We model the
power consumption of the AppOS task, which comprises the target
sensing application, source-channel coding and transmission.

\begin{figure}[h] \centering
\includegraphics [width=0.40\textwidth] {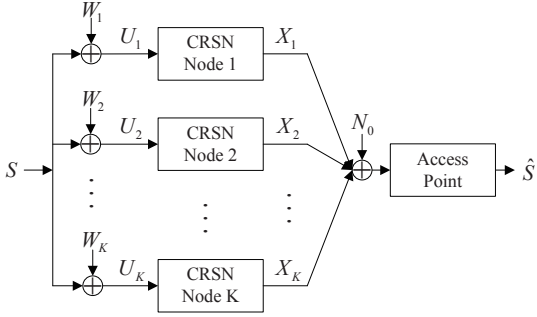}
\caption {Gaussian cognitive radio sensor network.} \label{Fig4}
\end{figure}

As shown in Fig.\ref{Fig4}, we consider the Gaussian source $S$ with
zero mean and variance $\sigma _S^2$. The source generates symbols
at a constant rate $L$ symbols per second. Every CRSN node's
observation includes a Gaussian noise $W_i$ with zero mean and equal
variance $\sigma _W^2$. The source $S$ is finally recovered as $\hat
S$ at the Access Point(AP).

%For a single node, the rate-distortion function \cite{R-D1} for
%source coding can be written in `bits per second' as:
%\begin{equation}
%R_{source} \left( D \right) = \frac{L}{2}\log _2 \left(
%{\frac{{\sigma _S^4 }}{{\left( {\sigma _S^2  + \sigma _W^2 }
%\right)D  - \sigma _S^2 \sigma _W^2 }}} \right) \label{eqn16}
%\end{equation}

%In multiple nodes scenario,

Every CRSN node first compresses its observation, then transmit it
to AP over the MAC using independently generated channel codes. This
is a multiterminal source coding system, and can be classified as
the CEO problem. For the symmetric Gaussian CEO problem, the $K$
nodes rate-distortion function \cite{R-D2} is:
\begin{equation}
R_{source} \left( D \right) = \frac{L}{2}\log _2 \left(
{\frac{{\left( {\frac{{\sigma _S^2 }}{D}} \right)^{\frac{1}{K}}
}}{{1 - \frac{{\sigma _W^2 }}{K}\left( {\frac{1}{D} -
\frac{1}{{\sigma _S^2 }}} \right)}}} \right) \label{eqn16}
\end{equation}

Note that we will only use the Gaussian source for illustration
later. For other sources, explicit form of the rate-distortion
function hasn't been derived. However, the outer bound can be
obtained, which is exactly in the form of (12) \cite{R-D2}. The
outer bound represents the worst case, which means for a given
source variance $\sigma _S^2$, the Gaussian sources are the most
difficult to compress.

We assume that the communication channel of interest is AWGN
channel. According to the Shannon Channel Capacity Theorem:
\begin{equation}
R_{channel}  \le W\log _2 \left\{ {1 + \frac{P}{{N_0 W}}} \right\}
\label{eqn17}
\end{equation}
where $W$ is the channel bandwidth, and $N_0$ is the unilateral
noise power spectral density. The energy for correctly delivering of
every bit of source information is:
\begin{equation}
E_{bit}  = \frac{P}{R} = N_0 W\frac{{2^{\frac{{R_{channel} }}{W}}  -
1}}{{R_{channel} }} \label{eqn18}
\end{equation}

Thus, the average AppOS power consumption can be expressed as:
\begin{equation}
\begin{array}{l}
 P_{_{AppOS}}  = p'_t E_{bit} R_{channel}  \\
  = p'_t N_0 W\left( {2^{\frac{{R_{channel} }}{W}}  - 1} \right) \\
 \end{array}\label{eqn19}
\end{equation}

We should point out that the source is encoded at rate $R_{source}$.
And $R_{source}$ is determined by the distortion $D$, the number of
nodes $K$, the variance of source $\sigma _S^2$ and the variance of
noise $\sigma _W^2$, regardless of the PU activity.

However, only a fraction of $p_t$ throughout the time domain can be
used for effective transmission. Therefore, in order to offset the
slots forbidden for transmission, the channel coding rate should be
higher than source coding rate:
\begin{equation}
\begin{array}{l}
 R_{channel}  = \frac{{R_{source} }}{{p_t }} \\
  = \frac{L}{{2p_t }}\log _2 \left( \frac{{\left( {\frac{{\sigma _S^2 }}{D}} \right)^{\frac{1}{K}} }}{{1 - \frac{{\sigma _W^2 }}{K}\left( {\frac{1}{D} - \frac{1}{{\sigma _S^2 }}} \right)}}
 \right) \\
 \end{array} \label{eqn20}
\end{equation}

From (8), (15) and (16), we formulate the average AppOS power with
respect to $p_t$:
\begin{equation}
\begin{array}{l}
 P_{_{AppOS}}  \left( {p_t} \right) = \left( {p_t  + P_E p({\rm H}_1 )} \right)N_0 W \\
  \times \left( {\left( \frac{{\left( {\frac{{\sigma _S^2 }}{D}} \right)^{\frac{1}{K}} }}{{1 - \frac{{\sigma _W^2 }}{K}\left( {\frac{1}{D} - \frac{1}{{\sigma _S^2 }}} \right)}}
 \right)^{\frac{L}{{2p_t W}}}  - 1} \right) \\
 \end{array} \label{eqn21}
\end{equation}

\emph{Proposition 1:} $P_{_{AppOS}} \left( {p_t}\right)$ is a
monotonically decreasing function.

\emph{Proof: See Proof of Proposition 1 in Appendix B}

The result can be confusing at the first glance, since we may
intuitively think that the AppOS power would grow with the
transmission probability. However, this is not the case. Now we
provide a heuristic understanding. If the transmission probability
$p_t$ is very low, the channel coding rate in the transmitting slots
has to be very high to make up for those silent slots. According to
(14), the transmission becomes less power efficient. Therefore, for
certain distortion and source coding rate, the average AppOS power
decreases with $p_t$.

\subsection{Joint Power Consumption Model}

On the one hand, if we allocate more power for AmOS, we are more
confident about the status of the primary user, therefore we can
grasp more opportunities for transmission. On the other hand,
delivering the information of the target source to the AP also
requires energy; the more power we allocate to AppOS, the higher
source and channel coding rate we can achieve. Under the condition
that power is constrained in CRSN node, we face a dilemma on how to
balance the two tasks. The effective transmission probability $p_t$
is the key parameter that naturally connects the two sensing tasks.

From (10) and (17), the total power consumption can be modeled as a
function of $p_t$:
\begin{equation}
\begin{array}{l}
 P_{total} \left( {p_t} \right) \\
  = \left( {\frac{{Q^{ - 1} \left( {1 - \frac{{p_t }}{{p({\rm H}_0 )}}} \right) - \sqrt {\frac{{2\sigma _s^2 }}{{\sigma _u^2 }} + 1} Q^{ - 1} \left( {1 - P_E } \right)}}{{\sigma _s^2 /\sigma _u^2 }}} \right)^2  \times \frac{{E_{sample} }}{T} +  \\
 \left( {p_t  + P_E p({\rm H}_1 )} \right)N_0 W\left( {\left( \frac{{\left( {\frac{{\sigma _S^2 }}{D}} \right)^{\frac{1}{K}} }}{{1 - \frac{{\sigma _W^2 }}{K}\left( {\frac{1}{D} - \frac{1}{{\sigma _S^2 }}} \right)}}
 \right)^{\frac{L}{{2p_t W}}}  - 1} \right) \\
 \end{array}\label{eqn22}
\end{equation}

\emph{Proposition 2:} When the probability of false alarm
$P_{FA}<\frac{1}{2}$, $P_{total} \left( {p_t} \right)$ is a convex
function with respect to $p_t$. That is to say, we can obtain the
minimal total power consumption and a unique power efficient
allocation solution for the CRSN node, if $P_{FA}$ falls into this
range.

\emph{Proof: See Proof of Proposition 2 in Appendix A}

\emph{Theorem 1:} Under our slotted sensing and transmission scheme,
there is always a minimal total power consumption and corresponding
optimal power allocation scheme for the CRSN to achieve certain
distortion constraint.

\emph{Proof: See Proof of Theorem 1 in Appendix B}

We end this section by summarizing the above results. In the cases
when $P_{FA}<\frac{1}{2}$, we know $P_{total} \left( {p_t} \right)$
is convex from \emph{Proposition 1}. We can thus design efficient
search algorithm to find the optimal power consumption. Otherwise,
\emph{Theorem 1} shows that, though the function is not convex, we
can still find the optimal power consumption through exhaustive
search, and calculate the corresponding power allocation scheme.

\section{Simulation Result}
To validate the analysis of the proposed energy efficient Joint
Source and Channel Sensing scheme, we present several numerical
results. We use Matlab as our simulator. For all scenarios, we set
the PU occupation rate to be $0.3$, which means the PU is active
with this probability. The max miss detection probability in
\textbf{AS4} is $0.1$; the energy consumed per sample in spectrum
sensing is $E_{sample}=0.1$mW; the source is of unit variance, i.e.
$\sigma _S^2  = 1$; the symbol rate of the source is $L = 1$M bauds;
the distortion is constrained to be 0.1. There are $K=10$ nodes and
the bandwidth of the considered AWGN channel is $W = 5$MHz.

\begin{figure}[h] \centering
\includegraphics [width=0.50\textwidth] {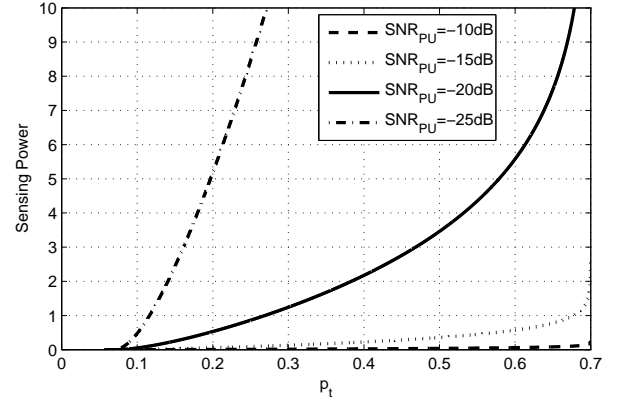}
\caption {Average AmOS power under different PU SNR.} \label{Fig5}
\end{figure}
\begin{figure}[h] \centering
\includegraphics [width=0.50\textwidth] {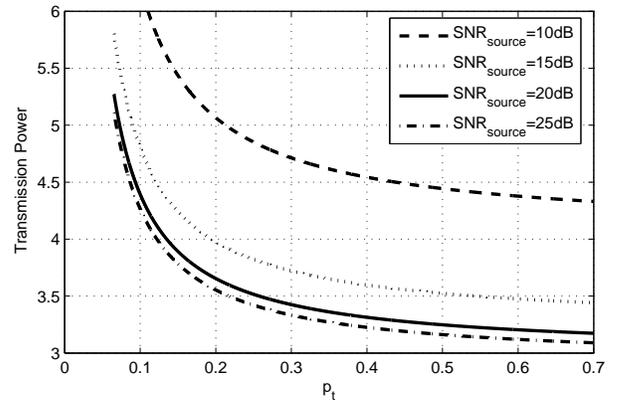}
\caption {Average AppOS power under different source SNR.}
\label{Fig6}
\end{figure}
\begin{figure}[h] \centering
\includegraphics [width=0.50\textwidth] {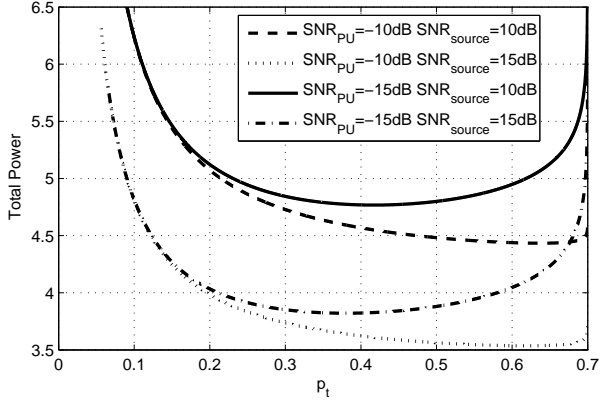}
\caption {Average total power under different PU SNR and source
SNR.} \label{Fig7}
\end{figure}

From Fig.5, we find that the average AmOS power increases with the
effective allowed transmission probability, i.e. the more we pay on
spectrum sensing, the more chances we obtain for transmission. We
can see from Fig.6 that the average AppOS power drops as
transmission probability increases, and this is consistent with the
analysis of \emph{Proposition 1}. Fig.5 and Fig.6 also show that as
the spectrum environment and monitored source become noisier, the
corresponding AmOS and AppOS power consumption increase.

Finally, Fig.7 shows that there is a unique valley point in every
curve, which corresponds to the optimal total power. Any other
transmission probability and power allocation scheme will result in
a higher total power consumption. In Fig.7, when the source SNR is
$10$dB and the PU SNR is $-15$dB, the optimal $p_t$ is $0.42$, and
the optimal total power is $4.8$W. $5.1\%$ of the power should be
allocated to AmOS to achieve optimality.

\section{Conclusion}
In this paper, we introduced a novel concept of Joint Source and
Channel Sensing for Cognitive Radio Sensor Networks, which seeks to
deliver the application source information to the access point in a
most power efficient manner. We presented a specific slotted sensing
and transmission scheme. By exploiting the relation between AmOS and
AppOS tasks, we modeled their power consumption properly and jointly
analyzed them. We proved that optimal power consumption and
corresponding power allocation scheme exist for fixed distortion
requirement. Finally, we present simulation results to support our
analysis.

\section{Appendix A: Proof of Proposition 2}
\emph{Proof:} The former part of (18) can be viewed as a composite
function $h\left( {p_t } \right){\rm{ }} = {\rm{ }}g\left( {f\left(
{p_t } \right)} \right)$, where $f\left( {p_t } \right) = Q^{ - 1}
\left( {1 - \frac{{p_t }}{{p({\rm{H}}_0 )}}} \right)$, and
\begin{equation}
g\left( x \right) = \left( {\frac{{x - \sqrt {\frac{{2\sigma _s^2
}}{{\sigma _u^2 }} + 1} Q^{ - 1} \left( {1 - P_E } \right)}}{{\sigma
_s^2 /\sigma _u^2 }}} \right)^2  \times \frac{{E_{sample} }}{T}
\end{equation}
Since$f\left( {p_t } \right) - \sqrt {\frac{{2\sigma _s^2 }}{{\sigma
_u^2 }} + 1} Q^{ - 1} \left( {1 - P_E } \right) = \sqrt N
\frac{{\sigma _s^2 }}{{\sigma _u^2 }} > 0$, and all other parameters
in $g(x)$ are non-negative, $g(x)$ is a convex and non-decreasing
function.

According to the property of inverse Q function, $f(p_t)$ is convex
as long as $p_t  > \frac{1}{2}p\left( {H_0 } \right)$, which is
equivalent to $P_{FA}  < \frac{1}{2}$.

Now that $f\left(  \cdot  \right)$ and $g\left(  \cdot  \right)$ are
convex functions and $g\left(  \cdot  \right)$ is non-decreasing,
then the former part $h(x) = g(f(x))$ is convex.

The latter part of (18) can prove to be convex through its second
order derivative:
\begin{equation}
\begin{array}{l}
 \frac{{\partial ^2 P_{_{AppOS}} }}{{\partial p_t ^2 }} =  \\
 \frac{{LN_0 \ln \left( C \right)C^{\frac{L}{{2p_t W}}} \left( {L\ln \left( C \right)\left( {p_t  + P_E p\left( {H_1 } \right)} \right) + 4p_t WP_E p\left( {H_1 } \right)} \right)}}{{4p_t^4 W}} \\
 \end{array}
\end{equation}
where $C = \frac{{\left( {\frac{{\sigma _S^2 }}{D}}
\right)^{\frac{1}{K}} }}{{1 - \frac{{\sigma _W^2 }}{K}\left(
{\frac{1}{D} - \frac{1}{{\sigma _S^2 }}} \right)}}
>1$, and all other parameters are positive. Obviously (20) is
positive, thus the latter part of (18) is also convex.

The sum power $P_{total} \left( {p_t} \right)$, as the sum of two
convex functions, is convex.

\section{Appendix B: Proof of Theorem 1}
\emph{Proof:} The derivative of $P_{_{AppOS}}$ is:
\begin{equation}
\begin{array}{l}
 P'_{_{AppOS}} \left( {p_t } \right) =  \\
  - \frac{{\left( {\left( {L\left( {1 + p\left( {H_1 } \right)} \right)\ln C - 2p_t W} \right)C^{\frac{L}{{2p_t W}}}  + 2p_t W} \right)N_0 }}{{2p_t }} \\
 \end{array}
\end{equation}

After observing (21), we can easily find that $P'_{_{AppOS}} \left(
{0^+}\right)=-\infty$ and $P'_{_{AppOS}} \left( {+\infty}\right)=0$.
Given that (24) is positive, we can conclude that $P'_{_{AppOS}}
\left( {p_t}\right)<0$, and $P_{_{AppOS}} \left( {p_t}\right)$ is
monotonically decreasing. Thus \emph{Proposition 1} is proved.

When $p_t$ falls in the range of (11),
\begin{equation}
\begin{array}{l}
 P_{_{AmOS}} ^\prime  \left( {p_t } \right) = \frac{{2E_{sample} }}{T}Q^{ - 1} \left( {1 - \frac{{p_t }}{{p({\rm{H}}_0 )}}} \right)^\prime   \times  \\
 \left( {\frac{{Q^{ - 1} \left( {1 - \frac{{p_t }}{{p({\rm{H}}_0 )}}} \right) - \sqrt {\frac{{2\sigma _s^2 }}{{\sigma _u^2 }} + 1} Q^{ - 1} \left( {1 - P_E } \right)}}{{\sigma _s^2 /\sigma _u^2 }}} \right) \\
 \end{array}
\end{equation}

It can be verified from (21) and (22) that
\begin{equation}
\begin{array}{l}
 P'_{total} \left( {0^ +  } \right) = P'_{_{AmOS}} \left( {0^ +  } \right) + P'_{_{AppOS}} \left( {0^ +  } \right) \\
  = 0 + \left( { - \infty } \right) =  - \infty  \\
 \end{array}
\end{equation}

Since $Q^{ - 1} \left( 0 \right)^\prime   =  + \infty $, we get:
\begin{equation}
\begin{array}{l}
 P'_{total} \left( {p\left( {H_0 } \right)^ -  } \right) \\
  = P'_{_{AmOS}} \left( {p\left( {H_0 } \right)^ -  } \right) + P'_{_{AppOS}} \left( {p\left( {H_0 } \right)^ -  } \right) =  + \infty  \\
 \end{array}
\end{equation}

(23) and (24) show that the continuous function $P_{total} \left(
{p_t} \right)$ decreases sharply at the left end and increases
sharply at the right end. Thus, there is a minimal total power
consumption point within the range of $p_t$, and we can calculate
the optimal $P_{_{AmOS}}$ and $P_{_{AppOS}}$ respectively.

\end{document}